\documentclass[12pt,letterpaper,top=2.5cm,bottom=2.5cm,left=3cm,right=2.5cm,marginparwidth=1.75cm]{article}
\usepackage{arxiv}

\usepackage{graphicx}

\usepackage{caption}
\usepackage{subcaption}
\usepackage{textcomp}
\usepackage{url}
\usepackage{colortbl}
\usepackage{soul}
\usepackage{multirow}
\usepackage{float}
\usepackage{physics}
\usepackage{pifont}
\usepackage{color}
\usepackage{alltt}
\usepackage[hidelinks]{hyperref}
\usepackage{enumerate}
\usepackage{siunitx}
\usepackage{breakurl}
\usepackage{epstopdf}
\usepackage{pbox}
\usepackage{stfloats}
\usepackage[vlined,linesnumberedhidden,ruled,resetcount]{algorithm2e}
\usepackage{algpseudocode}  
\usepackage{cite}
\usepackage{amsmath, amsfonts}

\newtheorem{remm}{Remark}

\begin{document}

\title{A Cyber-Physical Systems Framework for Tracking\\ Post Thermal-Runaway Temperature and Smoke Dynamics in Underground Mines}

\author{Yukta Pareek, 
	Khadija Omar Said, 
        Satadru Dey, 
	 and Ashish Ranjan Kumar 
    \thanks{Y. Pareek and S. Dey are with the Department of Mechanical Engineering, The Pennsylvania State University, University Park, 16802, USA (e-mail: ybp5153@psu.edu, skd5685@psu.edu).}%
		\thanks{K. O. Said and A. R. Kumar are with the Department of Energy and Mineral Engineering, The Pennsylvania State University, University Park, 16802, USA (e-mail: kos5600@psu.edu, awk5528@psu.edu).}
}

\maketitle

\maketitle
	
\begin{abstract}
Underground mining operations are actively exploring the use of large-format lithium-ion batteries (LIBs) to power their equipment. 
LIBs have high energy density, long cycle life, and favorable safety record.
They also have low noise, heat, and emission footprints. 
This fosters a conducive workplace environment for underground mining personnel. 
However, many occurrences of LIB failure have resulted in dangerous situations in underground mines.
The combustion products, including toxic emissions, can rapidly travel throughout the mine using the ventilation network. 
Therefore, it is critical to monitor the temperature and smoke concentration underground at all times to ensure the safety of the miners. 
High-fidelity models can be developed for specific scenarios of LIB failure, but are computationally prohibitive for large underground mine volumes, complex geometries, and long duration combustion events. 
To mitigate computation-related issues associated with high-fidelity models, we developed cyber-physical systems (CPS) models to examine temperature and smoke dynamics.
The mine supervisory control center, acting as the cyber framework, operates in conjunction with the physical underground mine.
The CPS models, trained on high-fidelity computational fluid dynamics (CFD) model data sets, present an exceptional estimate of the evolution of temperature and smoke concentration in the underground mine tunnel. 
Once implemented, the research results can help mine operators make informed decisions during emergencies.
\end{abstract}
 
\textbf{Keywords:}Lithium-ion battery, Thermal runaway, Underground Mines, Cyber-physical systems.

\section{Introduction}
\subsection{Motivation}
Underground mining operations are generally preferred when economics is unfavorable for the exploitation of minerals by surface mining methods. 
These operations play an indispensable role in meeting the growing demand for key metals, including several extracted from critical minerals, for a sustainable energy transition. 
Primarily, high-voltage or diesel-powered equipment is used for underground material transport.
Electricity-operated equipment drags cables along the mine floor, and diesel engines emit carcinogenic diesel particulate matter (DPM), creating hazardous conditions for underground mining personnel \cite{rawlins2023underground,weitekamp2020systematic}.
Therefore, following the prescribed regulatory framework, mining operations are investigating the use of large-format secondary lithium-ion batteries (LIBs) to power their equipment greatly because of their enhanced operational performance and flexibility. 
Their high energy density and Columbic efficiency, long cycle life, and low memory effect allow them to meet the requirements of high energy draining mining unit operations \cite{kim2019lithium}.

LIB systems have a strong safety record for a wide variety of applications when operated under their prescribed conditions. 
However, there have been several incidents in which they have explosively failed due to a swift combustion process called `thermal runaway' (TR) \cite{huang2021experimental, feng2015characterization}.
TR events are characterized by rapid irreversible exothermic reactions that result in fire, rapid increase in surrounding pressure fields, and explosive venting of flammable and toxic gases, including CO, CO$_2$, HF, NO, and NO$_2$ \cite{wang2019thermal, koch2018comprehensive,dubaniewicz2021thermal}. 
In underground mines, this could result from mechanical impact from a roof collapse, thermal stress, or electrical failures due to overcharging and discharging.
More LIB TR risks could arise from the most common emergencies in metal and non-metal underground mines, including fires and vehicular collisions \cite{de2019evaluation}.

Failure of a large-format LIB through TR could result in hazardous conditions for underground miners. 
Evidence can be observed in full-scale failure tests conducted on high-capacity LIBs.
It is important to note that extensive tests are restrictive due to the risky nature of TR, involving rapid emission of combustion products \cite{ouyang2025relationship,zou2024effects}. 
Therefore, mathematical simulations offer a viable substitute. Computational fluid dynamics (CFD) models have traditionally been the most promising of all numerical approximation methodologies.
Although they generate high-fidelity flow fields, they are computationally expensive, limiting their real-time deployment and requiring accurate boundary conditions and mesh resolution \cite{kong2022coupled}. 

LIBs are projected to become increasingly prevalent in underground mines to aid in the transition to automation and the phasing out of internal combustion engines, which are marked by relatively high heat loads, noxious emissions, and noise \cite{kuslap2025battery}.
Diesel particulate matter (DPM) emitted from these engines is a known carcinogen; its inhalation is known to negatively impact miners' health \cite{chang2017review}. 
Additionally, as mines move deeper into harsher conditions, LIB-powered equipment presents promising performance in terms of meeting sustainability targets, reducing occupational exposures to contaminants, and advancing towards fully automated zero-emission mining systems \cite{hooli2025battery}.
Despite this and their known TR hazards, limited studies have reported safety measures that can aid in a fast response.

\subsection{Related Works and Research Gaps}
Various mature methodologies, including calorimetry, are frequently used to study several aspects of LIB TR \cite{xu2020internal}. 
These techniques are mostly limited to laboratory-scale experiments. 
Therefore, these approaches are impractical for application in large volumes, such as underground mines that require large high-capacity LIBs \cite{hansen2010overview}.
Investigating the failure of these LIBs in confined environments necessitates the use of specialized numerical and experimental methods.
Previous research has included the development of mathematical superimposed models on heat exchange between smoke and tunnel walls that were validated with reduced-scale experiments \cite{wolski1995modeling}.
Three-dimensional CFD models have been extensively used to present fire events in underground facilities such as tunnels and roadways to study backlayering and other complex phenomena \cite{cheng2001simulation, fan2023prediction, yao2022theoretical}.
Fire dynamics simulator software (FDS) developed at the National Institute of Standards and Technology (NIST) has significantly advanced combustion modeling capabilities. 
FDS has been used to analyze temperature dynamics, smoke propagation, and visibility changes, in addition to evaluating the effectiveness of water sprinklers in various contexts, particularly within subterranean storage facilities and tunnels \cite{kallianiotis2022evaluation}.
It has been used to study combustion products such as CO\textsubscript{2} and CO generation in underground mine fires \cite{yuan2016modeling, salami2024enhancing, kumar2024comparative}.
However, despite their efficacy in presenting high-fidelity results, these models are computationally expensive owing to high grid resolution, require a precise understanding of the boundary conditions and flow domain, and therefore cannot be implemented quickly for dynamic mining conditions where combustion source, for example, may have variable locations and the flow volume might change frequently due to continuous extraction of minerals. 

Real-time fault detection is critical for large-format battery applications to enable immediate decision-making in underground mine emergencies. 
Several methods are used to assess thermal states and identify thermal faults in real time.
For example, research work has been performed to diagnose faults in LIB cells using the observer-based partial differential equations (PDE) method \cite{8166763}. 
A method was presented to estimate the 2D temperature distribution in LIB pouch cells using the optimal sensor placement and estimation algorithm. 
This research used a 2D cell configuration to present the non-uniform temperature profile \cite{9399498}.
Another research presented results from an electrochemical impedance spectroscopy-based method and performed a sensitivity analysis of the LIB impedance to temperature and state of charge. This work showed that appropriate estimation parameters could be obtained at low frequencies \cite{BEELEN2016128}. 
In another research, an extended Kalman filter-based approach was presented to estimate the surface and core temperature of a cell. Subsequently, a dual-extended Kalman filter using the same reduced-order thermal model was shown to estimate the convection coefficient \cite{7097077}. 
Deep neural network-based methods have also been used for surface temperature estimation. 
As an example, models using a feedforward neural network and a recurrent neural network using the long short-term memory framework were used to estimate the temperature with a root-mean-square error (RMSE) less than 2 \textdegree C \cite{9863003}. 

The aforementioned works focus on the estimation of surface and internal temperatures of LIBs.  
They do not explore the impact of thermal runaway on the environment. 
Accurate temperature and smoke estimation in underground mines is crucial for assessing tolerance and visibility during self-escape and mine rescue missions.
In addition, they provide information on the extent of toxicity and ventilation performance during fire incidents \cite{wang2024thermal,hao2024numerical}.
Various techniques have been used to estimate the smoke concentration and its dispersion in underground mines.
These include empirical and semi-empirical methods that investigate the dispersion of smoke from various sources, including conveyor belts and cables \cite{si2024experimental, wang2025study}.
Numerical methods, including CFD-based models developed using open-source and commercial codes such as FDS, ANSYS, OpenFOAM, and COMSOL, have been widely used to estimate smoke under various ventilation conditions \cite{chen2024study,addis2024enhancing, guo2025study}.
Although these methods are precise, they require a lot of resources and take a significant amount of time, which limits their real-time application.
On the other hand, sensor-based estimation methods inferring smoke concentration are emerging as promising tools in underground mines \cite{gopalakrishnan2024system,huang2024research}. 
Despite this development, limited studies have combined CFD models with cyber-physical systems for real-time smoke concentration estimation in underground mines. 
This presents a major avenue for contributing to the science of transient-state evolution of the subterranean atmosphere during a major combustion event.

\subsection{Main Contribution}
In light of the aforementioned gap, we present results from our research on the high-fidelity three-dimensional numerical models and reduced-order models for thermal-runaway-type failure of large-format lithium-ion batteries. 
We developed models for an underground tunnel that forms the fundamental building block of large underground mining operations.
The main contributions of this work are as follows: 
\emph{we present a real-time approach for tracking temperature and smoke concentration dynamics in underground mines -- leveraging reduced order models, real-time estimation theory, and a cyber-physical system (CPS) setup.} 
Specifically, (i) we start by understanding the post battery thermal-runaway effects utilizing CFD simulations. 
Subsequently, (ii) we develop reduced order models for coupled smoke and temperature evolution dynamics under battery fire -- and then (iii) identify the parameters of the reduced order models using CFD simulation data. 
Next, (iv) we formulate a CPS setup in which the physical system (the underground mine) is equipped with sensors placed along the mining tunnels. 
These sensors measure temperature and smoke concentration at known discrete locations in the tunnel. Transient-state smoke and temperature data sets are sent to the cyber part of the system, which is the mine supervisory control center. 
The mine supervisory control center hosts an estimation algorithm that receives the data sets and combines them with reduced-order models to track the spatio-temporal evolution of smoke concentration and temperature. 
The estimation algorithm has been developed using the moving horizon estimation framework \cite{rao2001constrained}.

\subsection{Summary of the Paper}
To alleviate the drawbacks of the high-fidelity CFD models and make them amenable to implementation for a wide range of combustion events, we propose the development of CPS for rapid spatio-temporal prediction of the combustion products. 
This paper summarizes our findings from the research using an ensemble of high-fidelity CFD models and a CPS framework. 
Section \ref{problem} describes the effects of post thermal runaway events using CFD. 
We present the real-time estimation framework in Section \ref{rte}.
The simulation results and the corresponding discussions are presented in Section \ref{sr}. Finally, we provide the research conclusions in Section \ref{con}. 


\section{Temperature and Smoke Dynamics Under Battery Thermal Runaway Event in Underground Mines} \label{problem}
 Integration of LIB-powered equipment into underground mining operations presents several benefits. 
In general, haulage equipment is a prime candidate for electrification due to its considerable flexibility. 
However, despite recent advances in battery materials, safety features, and battery management systems, TR continues to remain a hazard \cite{said2025perspective}. 
Large-format batteries, in particular, although a prerequisite to power large mining equipment, are significantly more hazardous during TR. 
In the unlikely event of their TR-driven failure, tracking the temperature rise and smoke concentration in underground mines becomes critical for many reasons.
Firstly, this information is crucial for timely mitigation of TR hazard using automated systems, including ventilation and fire suppression systems, to restore underground mines to safe conditions.
In addition, this information can inform on appropriate self-escape routes and mine rescue missions while avoiding intense build up of toxic and noxious gasses in addition to secondary catastrophes.
As such, this information is crucial to ensuring the safety of both mining workers underground and expensive mining assets \cite{onifade2021towards}. 

To investigate the impact of a large-format LIB TR, we developed high-fidelity CFD models to mimic the event. 
The models were developed for the geometry of an underground mine tunnel representative of an underground stone operation in the eastern United States. 
For this study, a section of a mine was considered, represented as a 9.0 m wide x 5.5 m tall tunnel. 
The tunnel was 80.0 m long and allowed for the operations of large haulage machinery.
An electric underground equipment, powered by a large-format LIB, measuring 3.0 m (width) x 10.0 m (length) x 1.0 m (height), was modeled as the fire source. 
The dimensions mimic a typical modern large haulage mining equipment used underground. 
The upper and lateral surfaces were the sites of active combustion. 
Parameters of interest, including temperature and smoke, were tracked to quantify combustion byproducts and their dispersion within the ventilation flow.
These probes collected data throughout the TR event, starting from normal battery conditions, through initiation and propagation, and extending into the post-TR decay phase, until temperature and smoke concentrations returned to baseline conditions.

\subsection{Setting up the FDS models}
NIST-developed FDS software was used to develop combustion models \cite{mcgrattan2013fire}. 
One of the key features of the code is its ability to run codes in parallel on a workstation using distributed compute cores.
It is a thermo-fluid code that uses the large-eddy simulation (LES) framework to present high-resolution data on smoke transport and heat generation from fires.  
This is achieved by resolving the dominant, energy-containing turbulent eddies in space and time while modeling the smaller subgrid-scale motions, enabling depiction of transient flows.
FDS works by solving a form of the Navier-Stokes equations using rectilinear grids, which is represented by the meshes. 
Eq. \ref{MBE} shows the total mass in the control volume, where,  
$\rho$ (kg/m$^3$) is the density of the fluid, and the time is represented by
$\mathbf{t}$ (s), $\mathbf{u}$  (m/s) is the velocity of fluid flow, and $\mathbf{\dot{m}_b'''}$ presents the addition of mass from small objects that cannot be resolved.
Eq. \ref{CE} shows conservation of momentum in the flow domain. 
The term $ \frac{\partial \mathbf{u}}{\partial t}$ represents the local acceleration of the fluid, which is the change in velocity of the fluid at a fixed point in space. 
The term $ \mathbf{u} \times \boldsymbol{\omega}$ represents the stretching of the vortex, which indicates the deformation of the vortices due to their interaction with the gradients in the fluid flow. 
The next two terms show the pressure contributions. 
The forces resulting from buoyancy, body contribution, and viscosity are present in the equation.
The simulation results are visualized using the associated tool called `SmokeView'.

\begin{subequations}
\begin{align}
    &\frac{\partial \rho}{\partial t} + \nabla \cdot (\rho \mathbf{u}) = \dot{m}_b'''\label{MBE} \\ 
    &\frac{\partial \mathbf{u}}{\partial t} - \mathbf{u} \times \boldsymbol{\omega} + \nabla H - \tilde{p}\nabla (1/\rho) = \frac{1}{\rho}[(\rho - \rho_0)\mathbf{g} + \mathbf{f}_b + \nabla \cdot \tau] \label{CE}\\
    &\frac{\partial {(\rho e)}}{\partial t}
     +(\rho e+p)\frac{\partial {u_i}}{\partial {x_i}} = \frac{\partial {(\tau_{ij} u_j)}}{\partial x_i}+\rho f_i u_i+\frac{\partial {(\dot{q}_i)}}{\partial x_i}+r 
    \label{TE}
\end{align}
\end{subequations}

The heat release rate (HRR) is one of the critical input parameters for the CFD models. 
The high  HRR values observed during large-format LIB-TR result in hazardous underground conditions, leading to the onset of secondary fires \cite{chen2020experimental, liu2021experimental, peng2020new}.
The dimensions of the grid cells required for the CFD simulations are calculated using the characteristics fire diameter, $\mathbf{D^*}$, given by Eq. \ref{eq:Dstar}. 
Here, $\mathbf{Q}$ is the heat release rate (HRR) (kW), the density of the ambient air is $\mathbf{\rho_\infty}$ (kg/m$^3$), 
the specific heat capacity of the air is $\mathbf{c_p} = $ 1.005 kJ/(kg$\cdot$K), the ambient temperature is represented by $\mathbf{T_\infty}$ (K), and $\mathbf{g} = 9.8\, m/s^2$ is the constant acceleration due to gravity. 
The goal of the $\mathbf{D^*}$ calculation is to ensure that the mesh is fine enough to resolve the dynamics of fire plumes.
For these sets of simulations, a coarse mesh was developed using the value $\mathbf{D^*}$ to 
(i) minimize computational cost, and (ii) because the predicted trends in transient-state temperature and smoke concentration were found to be acceptable for analysis using state-space models. 

\begin{equation}
D^* = \left( \frac{Q}{\rho_\infty \, c_p \, T_\infty \, \sqrt{g}} \right)^{\frac{2}{5}} 
\label{eq:Dstar}
\end{equation}

The entry and exit were assigned to the tunnel using \texttt{\&VENT} function.
A mechanical ventilation system was assumed to provide air at the entrance of the tunnel at a constant velocity of \texttt{1.0 m$\slash$s}.
The exit of the tunnel was left open. This setup enables the CFD model to mimic the tunnel connection to the ambient atmosphere, which allows a nominal air flow along with the removal of contaminants.

\subsection{Transient-state combustion modeling}
\begin{figure}[b!]
    \centering    
    \includegraphics[width=0.45\textwidth]{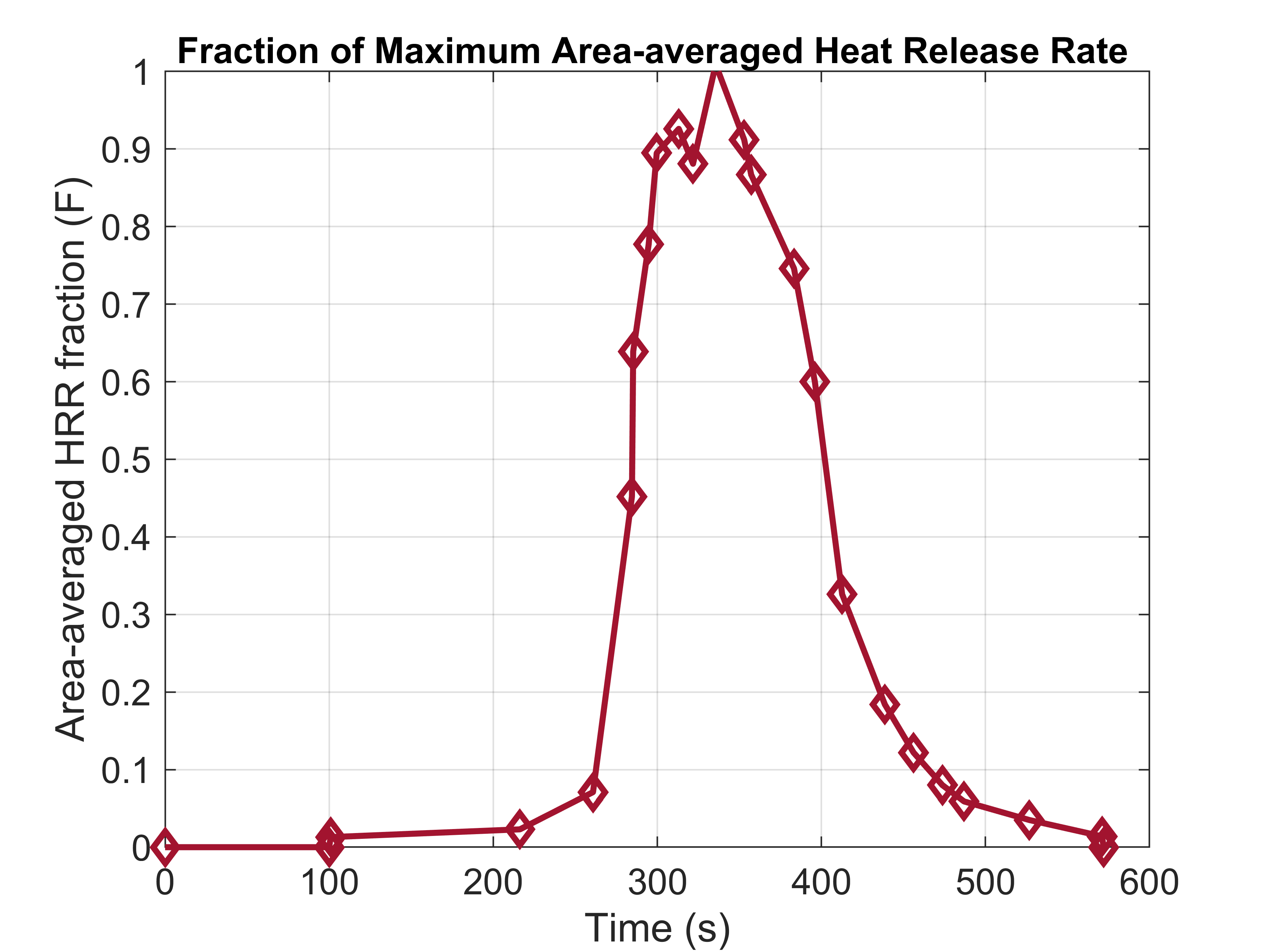}
    \caption{\small Plot of heat release rate (HRR) used as an input to the FDS models; heat released from the fire increases rapidly, which is then followed by a decay  \cite{peng2020new}.}
    \label{fig:HRR}
\end{figure}

The fire source, the LIB, was modeled using the \texttt{\&OBSTRUCTION} function to mimic a physical object that occupies a volume.
The heat injected into the flow domain is transferred through convection, radiative feedback, and conduction between the LIB and the tunnel.
This was modeled in the spatial bounds of  \( x=3.0\text{–}6.0~\mathrm{m},\ y=0.0\text{–}10.0~\mathrm{m},\ \text{and}\ z=0.5\text{–}1.5~\mathrm{m} \). 
The fire and inert zones of the LIB were defined using \texttt{\&FIRE SURFACE} and \texttt{\&INERT} respectively, to explicitly distinguish the actively burning surface releasing heat and species from non-reactive surfaces.
This is crucial for applying appropriate heat flux and emission rates.
The electrolyte was considered to be the main fuel as it is the most volatile part of the LIB with the lowest flashing point.
This is due to the rapid vaporization and exothermic oxidation after the decomposition temperature is exceeded. 
The HRR profile of the 243 Ah LIB was defined using the \texttt{\&RAMP} function, indicating the fractional progress of the peak HRR with a monotonic increase in time (Fig. \ref{fig:HRR}).
The temperature and smoke probes were modeled in the center of the tunnel at 4.5 m from the side walls and 1.5 m high.
These were modeled from the entrance and, after every 8.0 m, equidistant throughout the tunnel.
The simulations were run on a high-performance computing system. 

\begin{figure}[b!]
    \centering
    \includegraphics[width=0.49\textwidth]{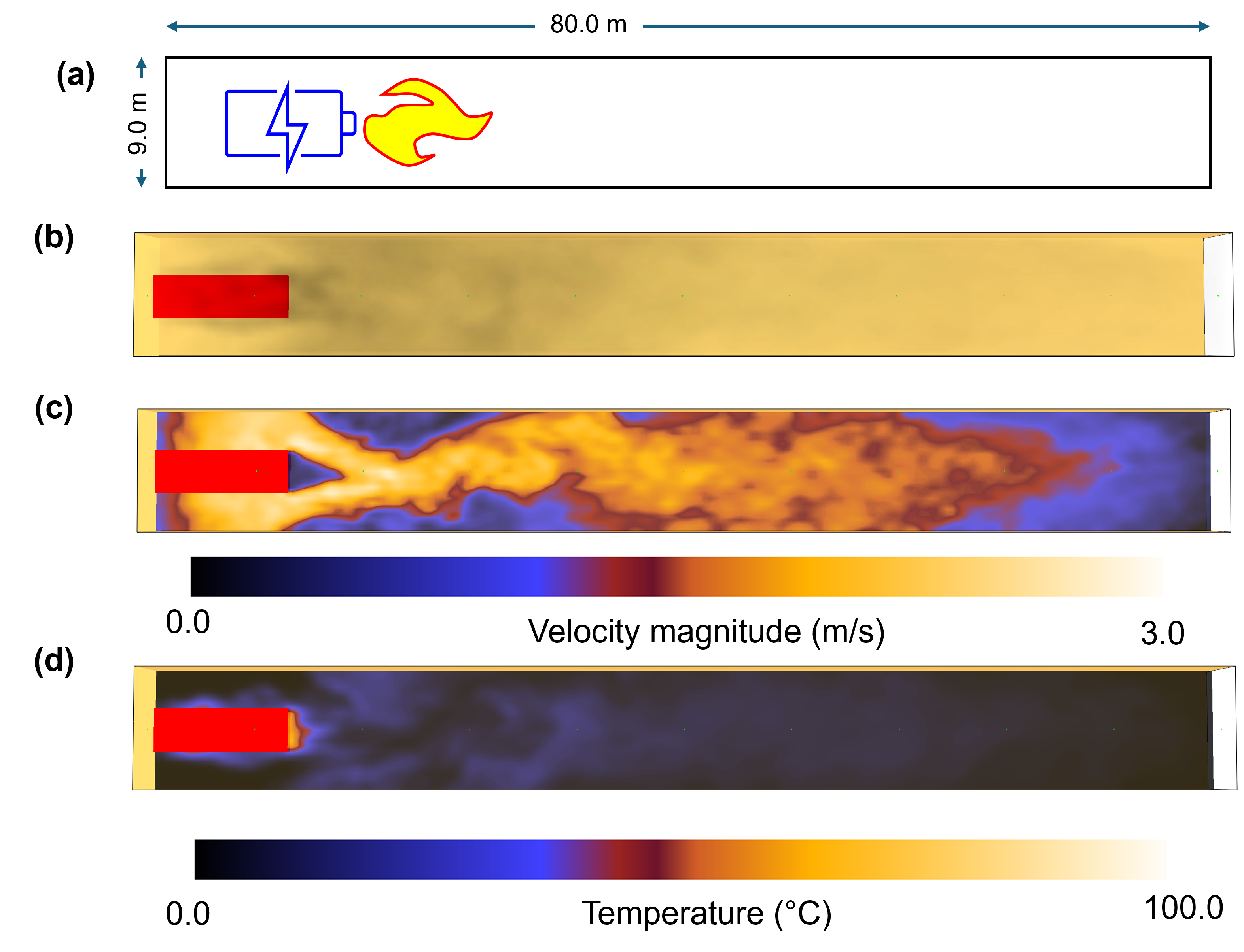}
    \caption{\small CFD modeling of the combustion event: (a) Plan view of the 80.0 m long mine tunnel, (b) Smoke concentration in the mine at time, $t = 150.0 s$, (c)  Contours of velocity magnitude; air is accelerated around the combustion sources due to constricted cross section and high temperature, and (d) high temperature near the battery.}
    \label{fig:smoke1}
\end{figure}

\begin{figure}[h!]
    \centering
    \includegraphics[width=0.49\textwidth]{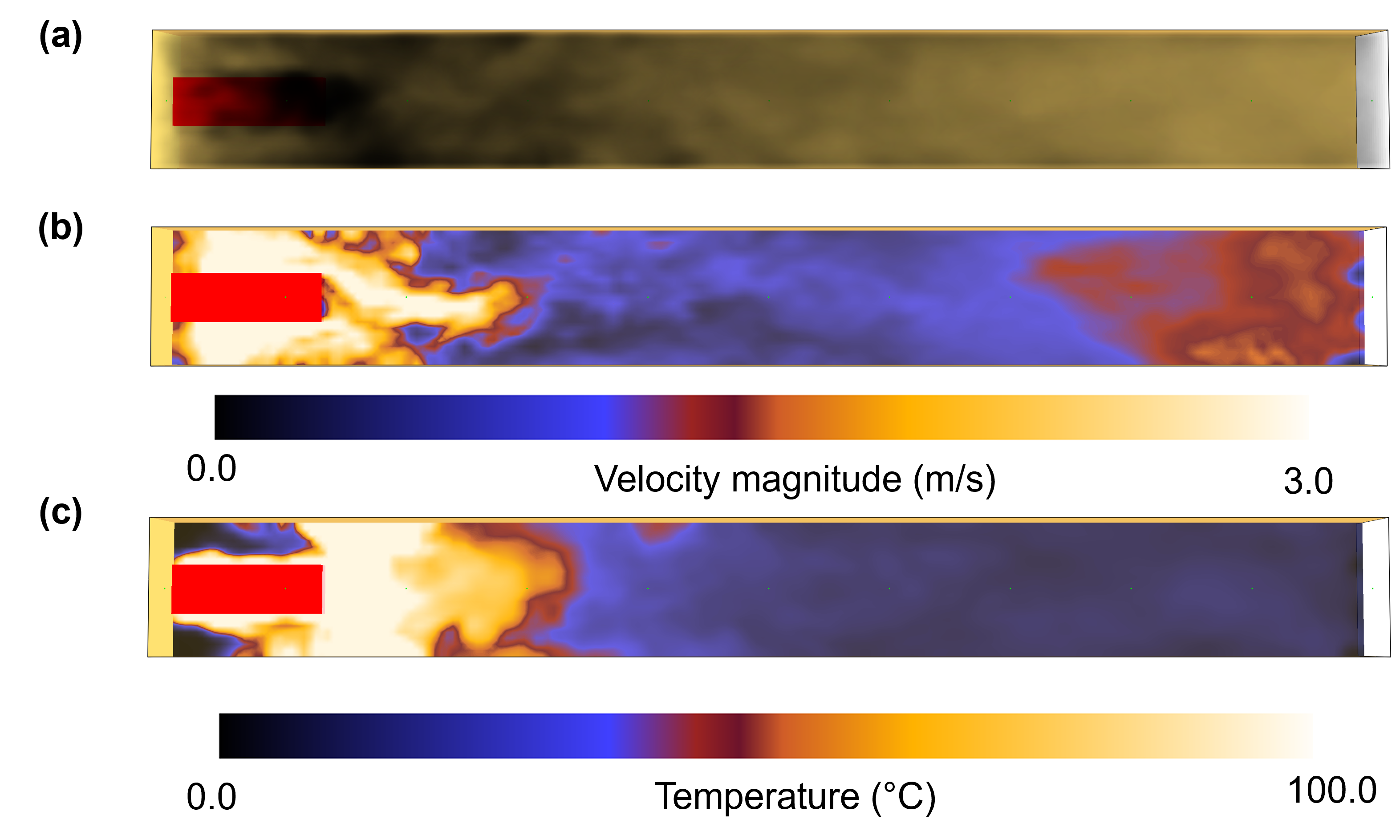}
    \caption{\small CFD modeling of the combustion event: (a) Plan view of the 80.0 m long mine tunnel, (b) Smoke concentration in the mine at time, $t = 275.0 s$, (c)  Contours of velocity magnitude; air is accelerated around the combustion sources due to constricted cross section and high temperature, and (d) high temperature near the battery.}
    \label{fig:smoke2}
\end{figure}

\begin{figure}[h!]
    \centering
    \includegraphics[width=0.49\textwidth]{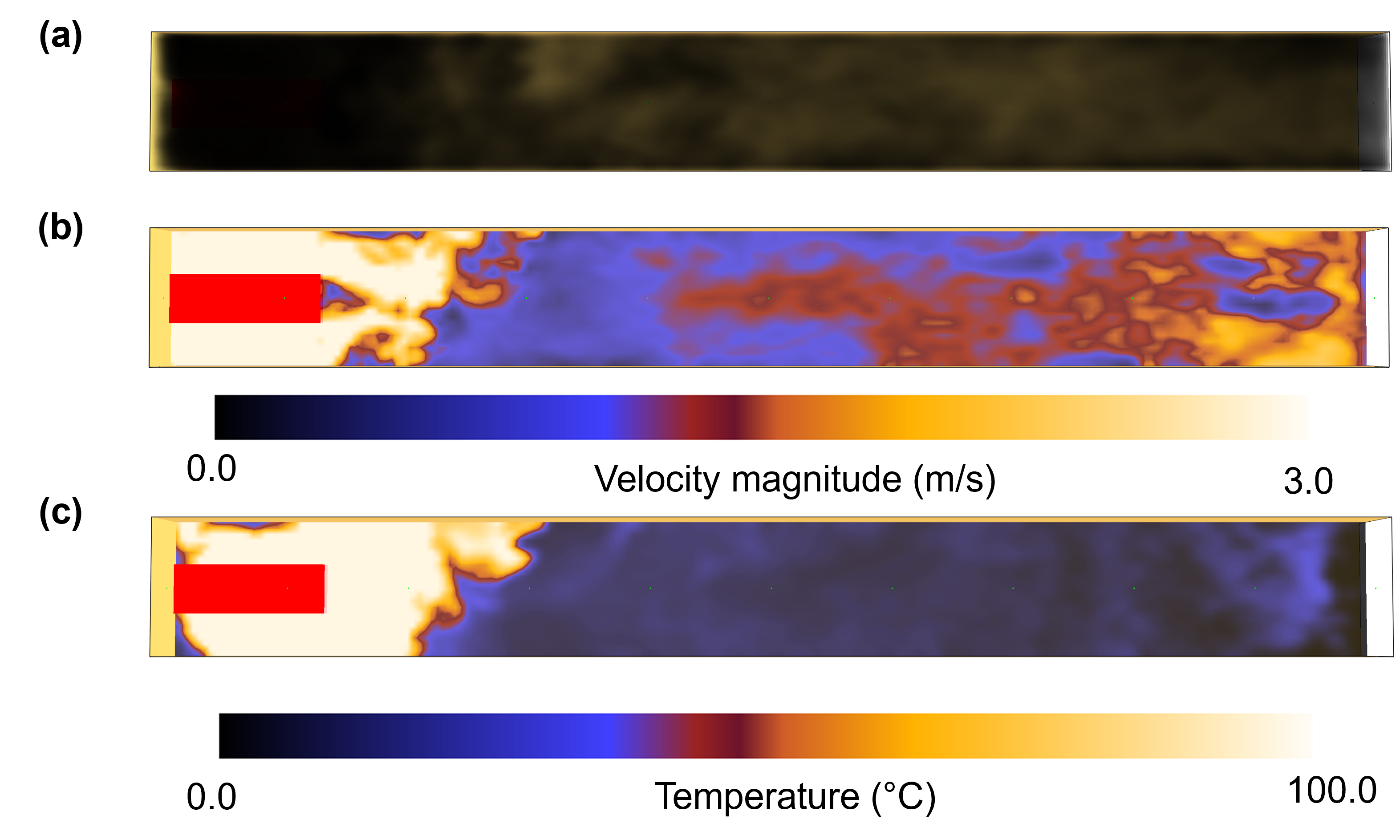}
    \caption{\small CFD modeling of the combustion event: (a) Plan view of the 80.0 m long mine tunnel, (b) Smoke concentration in the mine at time, $t = 400.0 s$, (c)  Contours of velocity magnitude; air is accelerated around the combustion sources due to constricted cross section and high temperature, and (d) high temperature near the battery.}
    \label{fig:smoke3}
\end{figure}

\begin{figure*}[h]
    \centering    
    \includegraphics[width=\textwidth]{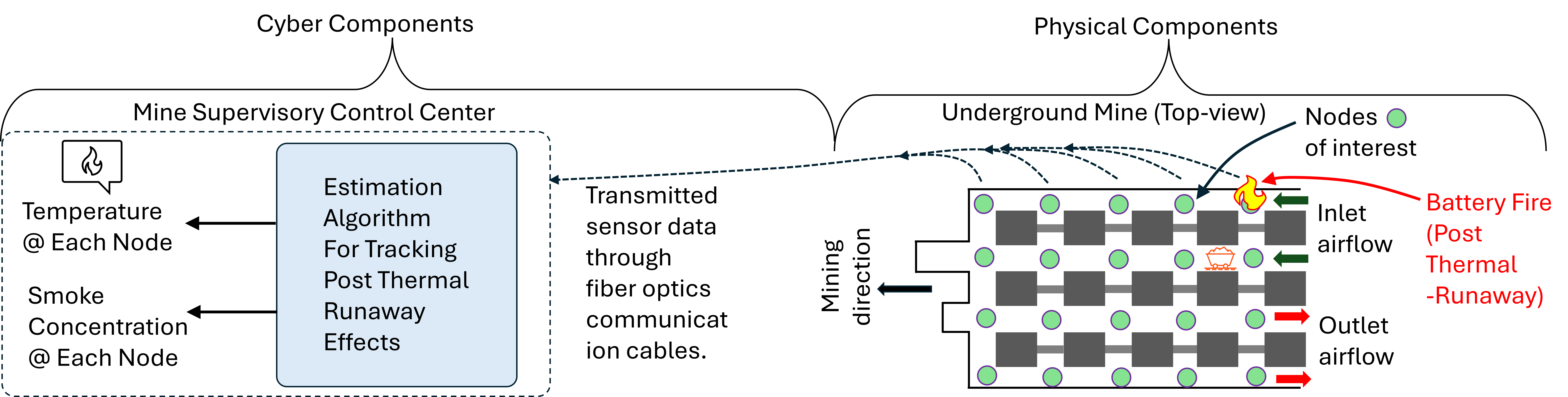}
    \caption{\small Cyber-Physical framework for tracking post thermal-runaway events in underground mines.}
    \label{fig:model}
\end{figure*}

The files intended for post-processing were saved at regular intervals.
These files contained information on the thermo-fluid field in the tunnel. 
Temperature and smoke concentrations were saved at the unique nodes every second. 
It is observed that the fire growth followed the heat release rate of the LIBs.
As the combustion event progresses, the HRR increases until it reaches its peak value.
Fig. \ref{fig:smoke1} shows the schematic of the setup of the model.
LIB was installed near the inlet and the tunnel was ventilated using airflow moving to the right as indicated by the direction of the flame. 
The composite figure also shows the smoke concentration, velocity magnitude (truncated to 3.0 m/s and), and temperature (on the scale of 0-100 \textdegree C) at time,  $t = 150.0\,s$. 
Similarly, Fig. \ref{fig:smoke2} shows the same parameters at time,  $t = 275.0\,s$. 
This is near the peak HRR. It is evident that intense HRR also affects the magnitude of the velocity as hot air is forced toward the ceiling of the tunnel.
Fig. \ref{fig:smoke3} shows the contours at $t = 400.0\,s$. This is also at a high HRR but towards a declining intensity.  
In all three images, high velocity is observed in the LIB because of a constricted cross section as well as heat injection into the system. 
The velocity magnitude contours also show the thermal driven flow profile. 
It is clearly observed that the contours of the high velocity magnitude move upwards due to increased heat injection as shown by the contours corresponding to time, $t = 275.0\,s$ and  $t = 400.0\,s$.
Similarly, as the HRR increases, smoke concentration also increases and decreases once the HRR peak is reached.
This is presented using the changing shades of black in the images.
It is also important to note that while the velocity magnitude and temperature contours are plotted on a plane, the smoke concentration is shown in the entire flow domain.


The spatio-temporal data generated from FDS simulations, including temperature and smoke concentrations, were extracted as a dynamic system response to the LIB TR in the tunnel.
Analysis of the data indicated important phases during TR, including an initial rapid escalation driven by exothermic energy release from LIB TR, a peak plume marked by maximum temperature and smoke concentrations, and gradual decay due to ventilation effects. 
Although crucial in explaining how temperature and smoke concentrations vary in time and space under ventilation conditions, these models are computationally expensive, limiting their application for accurate prediction and estimation.
As such, these data were used to formulate reduced-order state space models in which temperature and smoke concentrations served as measurable output in response to ventilation conditions. 
The resultant continuous time state space models allows timely assessment of TR evolution in real-time.
Ultimately, this data has been used to design state estimators that can infer unmeasured internal states including temperature and smoke concentrations.
Furthermore, these models will support development of anomaly detection and control algorithms.

\section{CPS-Based Framework for Real-Time Tracking of Temperature and Smoke Evolution} \label{rte}

The primary challenge in post-thermal runaway safety monitoring is the prohibitive computational cost of high-fidelity CFD models. 
Although these models provide invaluable insight into complex phenomena such as backlayering and turbulent smoke transport, their execution time renders them unsuitable for real-time decision support in an active emergency in an underground mining operation. 
To bridge this gap between high accuracy and real-time computational capability, we developed a cyber-physical systems (CPS) framework, as shown in Fig. \ref{fig:model}.
This framework leverages a data-driven reduced-order model that can be executed in a faster manner on standard industrial computing hardware, and augments such a model with a state estimation algorithm that reconstructs the full spatial profile of temperature and smoke concentration from a limited number of sensor measurements. This section details the development of this algorithm, from the system identification process to the design of the robust state estimator.

\begin{figure}[h!]
    \centering
    \includegraphics[width=0.7\textwidth]{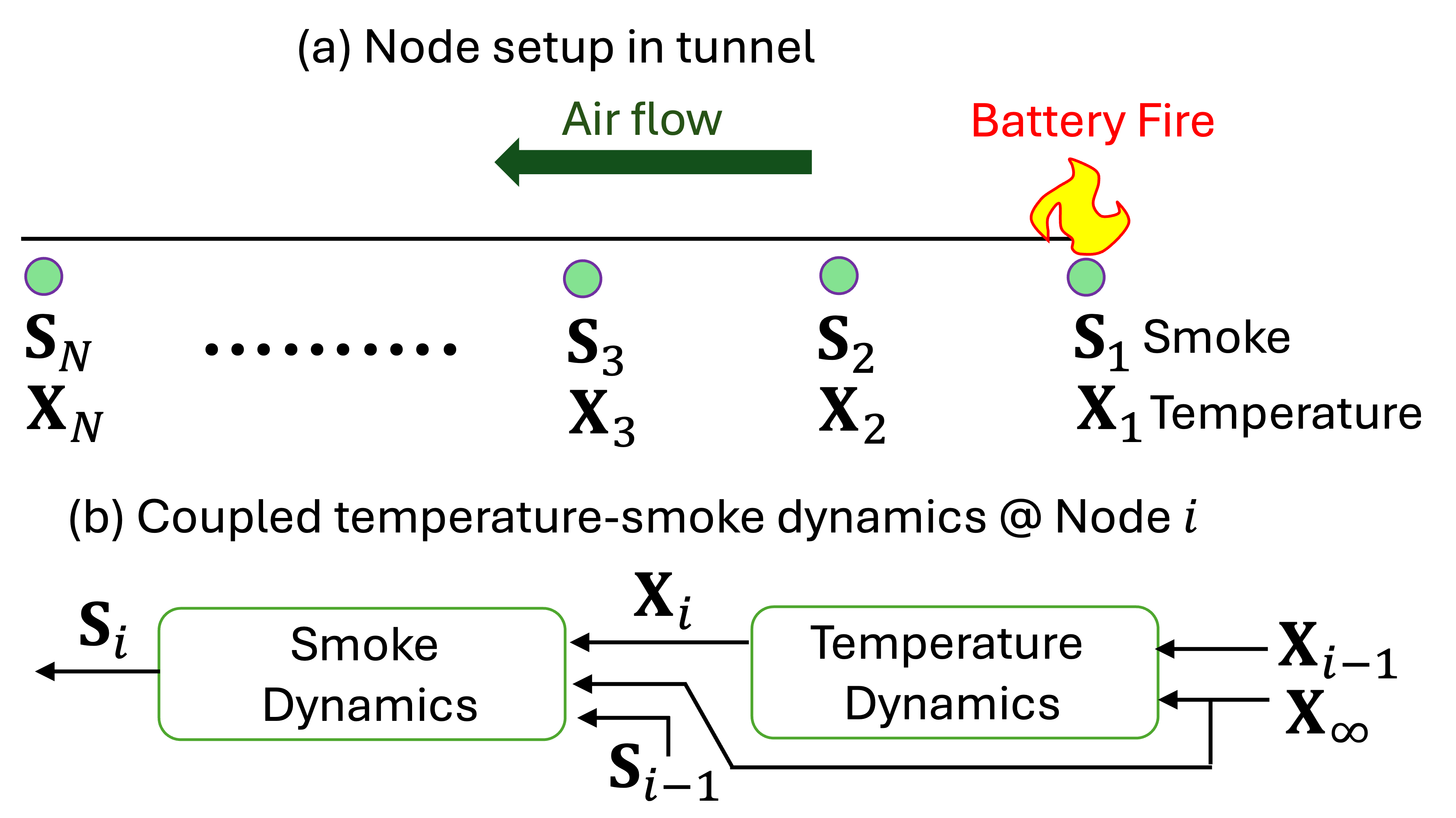}
    \caption{\small (a): A schematic of node setup in tunnel. (b): A sequential approach for modeling thermal and smoke dynamics in tunnel}
    \label{intro-11}
\end{figure}

\subsection{Data-driven state-space model development for temperature and smoke dynamics}
The high-fidelity CFD data, which capture the complex, nonlinear coupled dynamics of thermal and smoke propagation, were used to identify a reduced-order, linear time-invariant (LTI) state-space model suitable for real-time estimation, following an approach similar to our previous work \cite{said2025dynamical}. 
The spatial direction of the tunnel under consideration was discretized into $N$ nodes (as shown in Fig. \ref{intro-11} (a)). 
Subsequently, for each node in the tunnel state-space models were identified for the thermal and smoke dynamics;  
transforming the infinite-dimensional system into a tractable, lumped-parameter model.
This approach assumes that the coupling between temperature and smoke concentration is a secondary effect that can be captured implicitly through the model's input channels (as shown in Fig. \ref{intro-11} (b)). 
Next, we discuss the reduced order dynamical models for temperature and smoke propagation.

 \textbf{\textit{Thermal Dynamics Model}}: For node $i$ (where $i=1,\cdots,N$, as shown in Fig. \ref{intro-11} (a)), the continuous-time state-space representation for the thermal dynamics is given by Eq. \ref{state eqn-thermal} and \ref{output-eqn-thermal} shown below:

\begin{align}
    \dot{\mathbf{X}_i}(t) = \mathbf{A_T}_i\mathbf{X}_i(t) + \mathbf{B_T}_i\mathbf{u_T}_i(t), \label{state eqn-thermal}\\
    \mathbf{y_T}_i(t) = \mathbf{C_T}_i\mathbf{X}_i(t) + \mathbf{D_T}_i\mathbf{u_T}_i(t), \label{output-eqn-thermal}
\end{align}

where $\mathbf{X}_i(t)$ is the internal temperature-related state at node $i$; $\mathbf{u_T}_i(t)$ is the input vector given by $[{Q}(t), X_{\infty}(t)]^T$ for the node(s) where battery fire is present and $[X_{i-1}(t), X_{\infty}(t)]^T$ for the node(s) where no fire is present, where $Q$ and $X_{\infty}$ are the HRR and the ambient temperature, respectively; $\mathbf{y_T}_i(t)$ is the actual temperature of node $i$; the matrices $\mathbf{A_T}_i,\mathbf{B_T}_i,\mathbf{C_T}_i$ and $\mathbf{D_T}_i$ represent the state-space model parameters which are to be identified from the data. 

 \textbf{\textit{Smoke concentration dynamics model}}: For node $i$ (where $i=1,\cdots,N$, as shown in Fig. \ref{intro-11} (a)), the continuous-time state-space representation for smoke dynamics is given by Eq. \ref{state-eqn-smoke} and \ref{output-eqn-smoke}:
\begin{align}
    \dot{\mathbf{S}}_i(t) = \mathbf{A_s}_i\mathbf{S}_i(t) + \mathbf{B_s}_i\mathbf{u_s}_i(t), \label{state-eqn-smoke}\\
    \mathbf{y_S}_i(t) = \mathbf{C_s}_i\mathbf{S}_i(t) + \mathbf{D_s}_i\mathbf{u_s}_i(t) \label{output-eqn-smoke},
\end{align}
where $\mathbf{S}_i(t)$ is the internal smoke concentration-related state at the node $i$; $\mathbf{u_s}_i(t) = [{Q}(t), X_i(t), X_{\infty}(t)]^T$ is the input vector for the node(s) where battery fire is present and $\mathbf{u_s}_i(t) = [{S_{i-1}}(t), X_i(t), X_{\infty}(t)]^T$ for the node(s) where no fire is present; $\mathbf{y_S}_i(t)$ is the actual smoke concentration of node $i$; the matrices $\mathbf{A_s}_i,\mathbf{B_s}_i,\mathbf{C_s}_i$ and $\mathbf{D_s}_i$ represent the state-space model parameters which are to be identified from the data.


\begin{remm}
    {The state-space model for temperature and smoke depicts dynamic response to HRR from LIB TR.
    The HRR influences these parameters through thermal and buoyancy-driven flows and convection heat transfer.
    As HRR increases, it induces a rise in temperature due to accumulated thermal energy and simultaneously enhances smoke generation and upward movement through buoyant plumes.
    Higher temperatures reduce air density, generating upward flow that entrains and transports smoke. 
    Additionally, nodes closer to the LIB indicate higher temperature and smoke concentrations, unlike the nodes downstream due to heat loss to surrounding surfaces, entrainment of cooler air, and dilution effects. 
    }
\end{remm}

\subsection{Design of state estimation algorithm}
The state-space model developed in the previous subsection, while compact and computationally efficient, suffers from inaccuracies arising from its reduced order nature. In order to address this issue, state estimation algorithms couple the state-space model with feedback from measured sensor data. However, in the context of post-thermal runaway tracking, the full state vector is not directly measurable. 
That is, not every node has sensors installed to measure temperature and smoke concentration. 
This is because it is physically and economically infeasible to install and maintain temperature and smoke sensors at every discrete node along a mine tunnel.
As the active mining front progresses, the flow domain's dimensions and the relative position of the sensors change dynamically depending on the mine production rate and several other factors.
Sensors are typically placed only at critical or accessible locations, providing a sparse, low-dimensional measurement. The open-loop state-space model, which propagates the state based only on input $\mathbf{u}(t)$, is insufficient for safety-critical applications. Any discrepancy between the model and reality-caused by initial condition errors, unmodeled disturbances, or parameter inaccuracies-will cause the estimated state to diverge from the true state without correction. 

In this paper, we employ the Moving Horizon Estimation (MHE)  framework to estimate the temperature and smoke concentration states \cite{rao2001constrained}. 
The MHE is an optimization-based estimation technique that recursively solves a least-squares problem over a sliding window of the most recent measurements. 
This approach explicitly handles state constraints (i.e., temperatures and smoke concentrations) and systematically accounts for process and measurement noise, providing robust and accurate state estimates. 
Taking into account the dynamics of temperature and smoke concentration for each node, given by Eq. \eqref{state eqn-thermal}-\eqref{output-eqn-thermal} and \eqref{state-eqn-smoke}-\eqref{output-eqn-smoke}, we can represent both the temperature and smoke concentration for all nodes in a compact state-space model as:
\begin{align}
    & \dot{z}(t) = A z(t) + B u(t), \label{ss-1}\\
    & y(t) = C z(t) + D u(t) \label{ss-2},
\end{align}
where $z$ represents the full state vector, that is, $[\mathbf{X}_1, \mathbf{X}_2,\cdots,\mathbf{X}_N]^T$ for the temperature dynamics and $[\mathbf{S}_1, \mathbf{S}_2,\cdots,\mathbf{S}_N]^T$ for the smoke concentration dynamics; $u$ represents the system input vector in terms of $Q$ and $\mathbf{X}_\infty$; and $y$ represents the nodes where temperature and smoke concentration sensors are placed.
Before designing the MHE, the observability of the system was verified by ensuring that the observability matrix $\mathcal{O} = [{C}^T, ({C}{A})^T, ..., ({C}{A}^{N-1})^T]^T$ had full rank $N$. 
Subsequently, the model Eq. \eqref{ss-1}-\eqref{ss-2} was written in a discrete-time format as shown in Eqs. \ref{discrete state eqn-thermal} and \ref{discrete output eqn-thermal}:
\begin{align}
    &{z}(k+1) = (I+\delta t {A}){z}(k) + \delta t {B}{u}(k), \label{discrete state eqn-thermal}\\
    &{y}(k) = {C}{z}(k) + {D}{u}(k), \label{discrete output eqn-thermal}
\end{align}
where $\delta t$ is the sample time and $k$ is the time index.

The core of the MHE approach is to compute the state estimate at the current time $k$ considering a sequence of horizon length $W$. 
This involves solving a constrained optimization problem that finds the most likely sequence of states that explains the observed measurements, while respecting the system dynamics and constraints. Referring to the state-space model Eqs. \eqref{discrete state eqn-thermal}-\eqref{discrete output eqn-thermal}, at each time-step $k$, the MHE algorithm computes the state trajectory $\{z(k-W+1), \dots, z(k)\}$ over a horizon of length $W$, by solving the following optimization problem (Eq. \ref{eqn mhe cost}):

\begin{align}
    &\min_{\{z(j)\}_{j=k-W+1}^k} J_R + J_Q + J_\lambda, \label{eqn mhe cost}
\end{align}
with 
\begin{align}
    & J_R = \sum_{j=k-W+1}^{k} \| y(j) - C z(j) - Du(j) \|_R^2,\\
    & J_Q = \sum_{j=k-W+1}^{k-1} \| z(j+1) - Az(j)- B u(j) \|_Q^2,\\
    & J_\lambda = \lambda \| z(k-W+1) - \hat{z}(k-W+1) \|^2,
\end{align}
subject to the system dynamics \eqref{discrete state eqn-thermal}-\eqref{discrete output eqn-thermal}. Here, $\| z \|_M^2 = z^\top M^{-1} z$ for $M\in \{R,Q\}$ with $Q$ and $R$ being the weighting matrices, 
and $\lambda > 0$ is a regularization parameter to prevent ill-conditioning. 
The optimal estimate at the current time-step $k$ is given by Eq. \ref{rp}.

\begin{align}
    \hat{z}(k) = \arg\min_{\{z\}_{i=k-W+1}^k} J
    \label{rp}
\end{align}
where $J=J_R + J_Q + J_\lambda$ as given in Eq. \eqref{eqn mhe cost}. An algorithmic description of MHE is given in Algorithm \ref{abcd}

\begin{algorithm}[h!]
\caption{Algorithm for Moving Horizon Estimation (MHE) for Thermal and Smoke Concentration State Estimation}
\label{abcd}

\KwIn{Horizon length $W$, weighting matrices $Q$, $R$, regularization parameter $\lambda$}
\KwOut{State estimates $\hat{X}(k)$, $\hat{S}(k)$ at each time step $k$}

Initialize $\hat{X}(0) \gets X(0)$, $\hat{S}(0) \gets S(0)$, $k \gets 0$;\\
Buffers: $\mathcal{Y}_t, \mathcal{U}_t, \mathcal{Y}_s, \mathcal{U}_s \gets \emptyset$;

\While{new measurements available}{
    $k \gets k + 1$;\\[2pt]

    \textbf{Step 1: Data Buffer Update}\\
    $\mathcal{Y}_t \gets \mathcal{Y}_t \cup \{y_t(k)\}$ (keep last $W$ elements);\\
    $\mathcal{U}_t \gets \mathcal{U}_t \cup \{u_t(k)\}$ (keep last $W$ elements);\\
    $\mathcal{Y}_s \gets \mathcal{Y}_s \cup \{y_s(k)\}$ (keep last $W$ elements);\\
    $\mathcal{U}_s \gets \mathcal{U}_s \cup \{u_s(k)\}$ (keep last $W$ elements);\\[4pt]

    \textbf{Step 2: Determine Effective Horizon}\\
    \eIf{$k < W$}{
        $W_{\text{eff}} \gets k$;\\
        $\lambda_{\text{eff}} \gets \lambda \cdot \dfrac{W}{k}$;
    }{
        $W_{\text{eff}} \gets W$;\\
        $\lambda_{\text{eff}} \gets \lambda$;
    }

    \textbf{Step 3: Estimate Temperature then Estimate Smoke Concentration}\\
   Thermal states $\{\hat{X}(k-W_{\text{eff}}+1), \dots, \hat{X}(k)\}$;\\
    Solve states $\{\hat{S}(k-W_{\text{eff}}+1), \dots, \hat{S}(k)\}$;\\

    \textbf{Step 4: Extract Current Estimates}\\
    $\hat{X}(k) \gets \hat{X}(k)$ (from thermal optimization solution); \\
    $\hat{S}(k) \gets \hat{S}(k)$ (from smoke optimization solution); \\

    \textbf{Step 5: Update Priors for Next Iteration}\\
    \If{$k \geq W$}{
        Store $\hat{X}(k-W+1)$, $\hat{S}(k-W+1)$ for next regularization term;\\
    \textbf{Output:} Current estimates $\hat{X}(k)$, $\hat{S}(k)$
    }
}
\end{algorithm}

\section{Simulation Results}
\label{sr}

In this section,we discuss the simulation results to illustrate the proposed approach. We start with identification of the reduced order models, and then explain the tracking performance of the estimation algorithm.

\subsection{Identification of reduced order model}
The reduced order models for thermal and smoke concentration dynamics was discussed in Section III.A. 
The identification of the open-loop model essentially boils down to finding the matrices $\mathbf{A_T}_i, \mathbf{B_T}_i, \mathbf{C_T}_i,\mathbf{D_T}_i$ and $\mathbf{A_s}_i, \mathbf{B_s}_i, \mathbf{C_s}_i,\mathbf{D_s}_i$ in \eqref{state eqn-thermal}-\eqref{output-eqn-smoke}. 
This identification process was conducted using the System Identification Toolbox in MATLAB \cite{matsys}. The N4SID (Numerical Subspace State-Space System Identification) algorithm was used, which constructs the matrices directly from input-output data via orthogonal projections, avoiding numerical conditioning issues common in large CFD generated datasets. 
The high-fidelity CFD data served as the training and validation dataset. 
For the  training data set, time-series data for LIB TR of two LIB with (60 Ah battery capacity and 15K ambient temperature) and (243 Ah battery capacity and 25 \textdegree C as ambient temperature) were concatenated to allow the different behaviors of LIB TR. 
For validation dataset: LIB with 60Ah capacity at $T_{amb} = 25$ \textdegree C was used. 
Here, we used a setup of 10 discrete equidistant nodes located along the tunnel, i.e., $N=10$ in reference to Fig. \ref{intro-11}.

\begin{figure}[h!]
    \centering
    \includegraphics[width=0.5\textwidth]{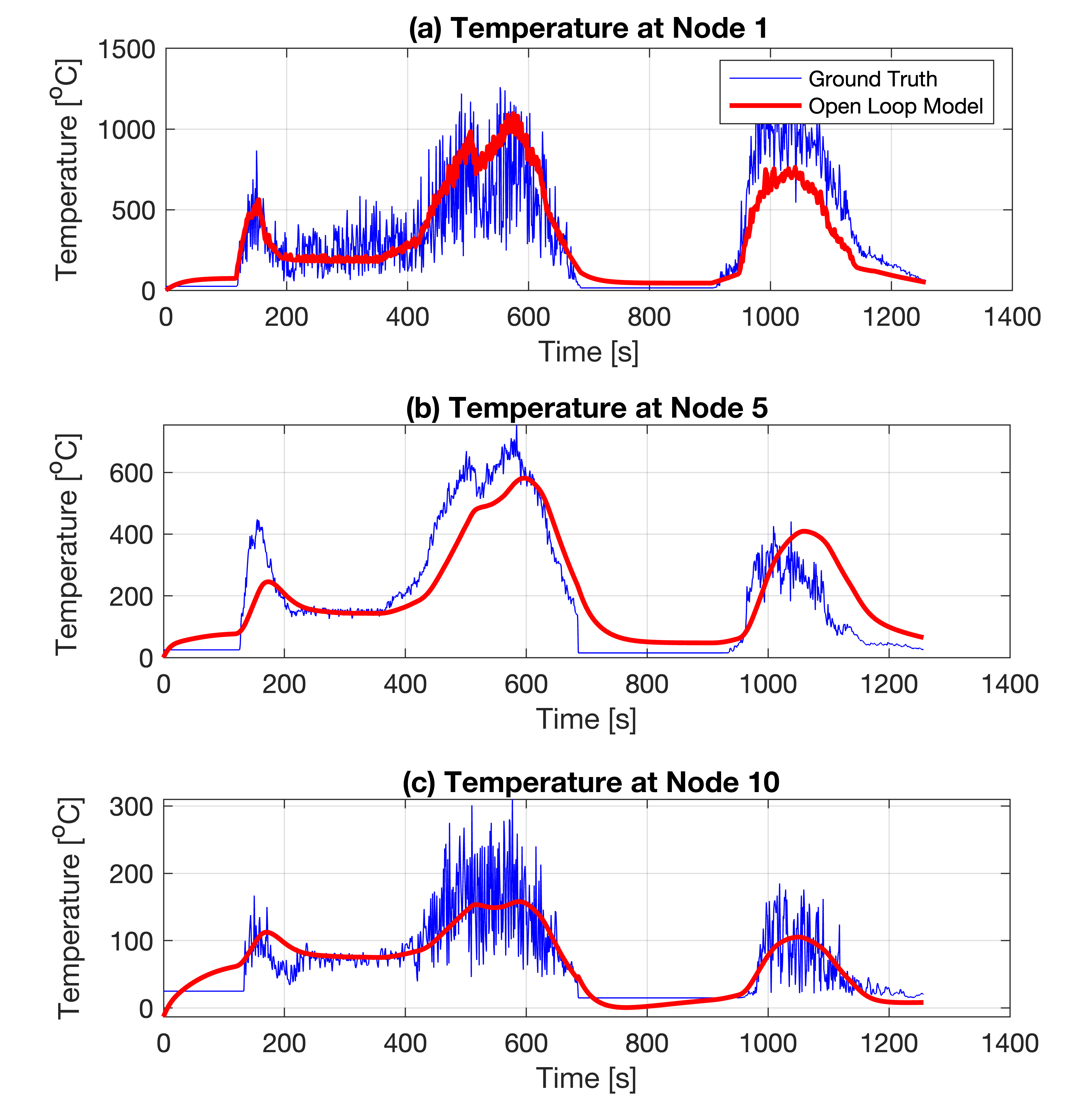}
    \caption{\small Comparison of reduced order thermal model and ground truth data for validation dataset. (a) Node 1, (b) Node 5, and (c) Node 10.}
    \label{OL_2_6_11_temp}
\end{figure}

\begin{figure}[h!]
    \centering
    \includegraphics[width=0.5\textwidth]{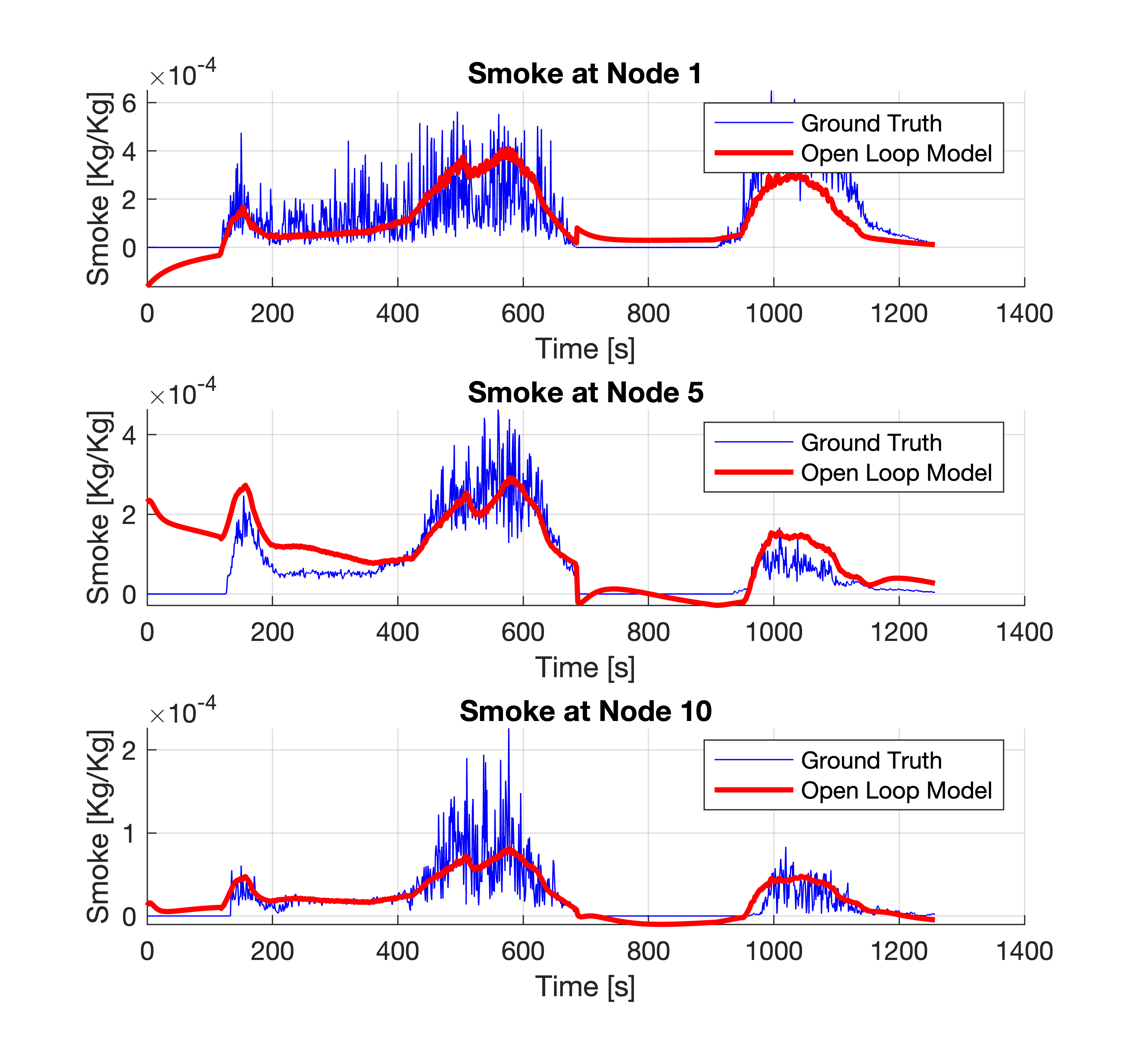}
    \caption{\small Comparison of reduced order smoke concentration model and ground truth data for validation dataset. (a) Node 1, (b) Node 5, and (c) Node 10.}
    \label{OL_2_5_11_smoke}
\end{figure}

Identification results for thermal model for a few nodes (nodes 1, 5, and 10) are shown in Fig. \ref{OL_2_6_11_temp}, which shows the temperature comparison between the reduced order model and the ground-truth temperature data. Similarly, Fig. \ref{OL_2_5_11_smoke} shows the comparison of smoke concentration between the reduced order model and ground truth data for nodes 1, 5, and 10. The root mean square errors (RMSE) between the reduced order models and the ground truth are given in Tables \ref{table_error_temp} and \ref{table_error_smoke}, which show reasonable accuracy of the reduced order models.

\begin{table}[h!]
    \centering
    \caption{\small Comparison of thermal open loop model RMSE and thermal estimation RMSE values for all 10 nodes.}
    \begin{tabular}{ | m{8em} | m{8em} | m{8em}|}
         \hline
         Node Number&
          Open Loop Model RMSE& 
         MHE RMSE \\ 
          \hline
           $N_1$ & 173.9791 & 0.4376\\ 
          \hline
           $N_2$ & 84.1008 & 190.3940 \\ 
           \hline
           $N_3$ & 95.5090 & 66.7118 \\ 
           \hline
           $N_4$ & 121.9038 & 34.8690 \\ 
           \hline
           $N_5$ & 94.7175 & 0.6079\\ 
           \hline
           $N_6$ &  79.5371& 36.2697 \\ 
           \hline
           $N_7$ &57.9318  & 56.5188\\ 
           \hline
           $N_8$ & 96.7051&  40.3930\\ 
           \hline
           $N_{9}$ & 52.2320 & 27.0042 \\ 
           \hline
           $N_{10}$ & 28.2565 & 0.0906 \\ 
           \hline
    \end{tabular}
    \label{table_error_temp}
\end{table}

\begin{table}[h!]
    \centering
    \caption{\small Comparison of smoke open loop model RMSE and smoke estimation RMSE values for all 10 nodes.}
    \begin{tabular}{ | m{8em} | m{8em} | m{8em}|}
         \hline
         Node Number&
          Open Loop Model RMSE& 
         MHE RMSE \\ 
          \hline
           $N_1$ & $1.0187 \times 10^{-4}$ & $2.6165\times 10^{-7}$\\ 
          \hline
           $N_2$ & $7.2876\times 10^{-5}$ & $1.1757\times 10^{-4}$ \\ 
           \hline
           $N_3$ & $4.2379\times 10^{-5}$ & $4.7887\times 10^{-5}$ \\ 
           \hline
           $N_4$ & $4.6170\times 10^{-5}$ & $5.1711\times 10^{-5}$ \\ 
           \hline
           $N_5$ & $7.1268\times 10^{-5}$ & $7.7231\times 10^{-6}$\\ 
           \hline
           $N_6$ &  $4.4228\times 10^{-5}$& $3.0887\times 10^{-5}$ \\ 
           \hline
           $N_7$ &$3.9120\times 10^{-5}$  & $2.8156\times 10^{-5}$\\ 
           \hline
           $N_8$ & $3.8651\times 10^{-5}$&  $3.0658\times 10^{-5}$\\ 
           \hline
           $N_{9}$ & $3.3825\times 10^{-5}$ & $2.6481\times 10^{-5}$ \\ 
           \hline
           $N_{10}$ & $1.5902\times 10^{-5}$ & $1.6746\times 10^{-6}$ \\ 
           \hline
    \end{tabular}
    \label{table_error_smoke}
\end{table}

\subsection{Tracking performance by moving horizon estimation algorithm}
In this subsection, we discuss the performance of the MHE algorithm for thermal and smoke concentration estimation, as discussed in Section III.B. We consider that there are only three nodes (namely, nodes 1, 5 and 10) where temperature and smoke concentration sensors are located. All the other nodes do not have any sensors installed. The goal of the estimation algorithm is to track the states in all nodes -- using the reduced order model and the feedback from the sensor data.

\begin{figure}[h!]
    \centering    \includegraphics[width=0.5\textwidth]{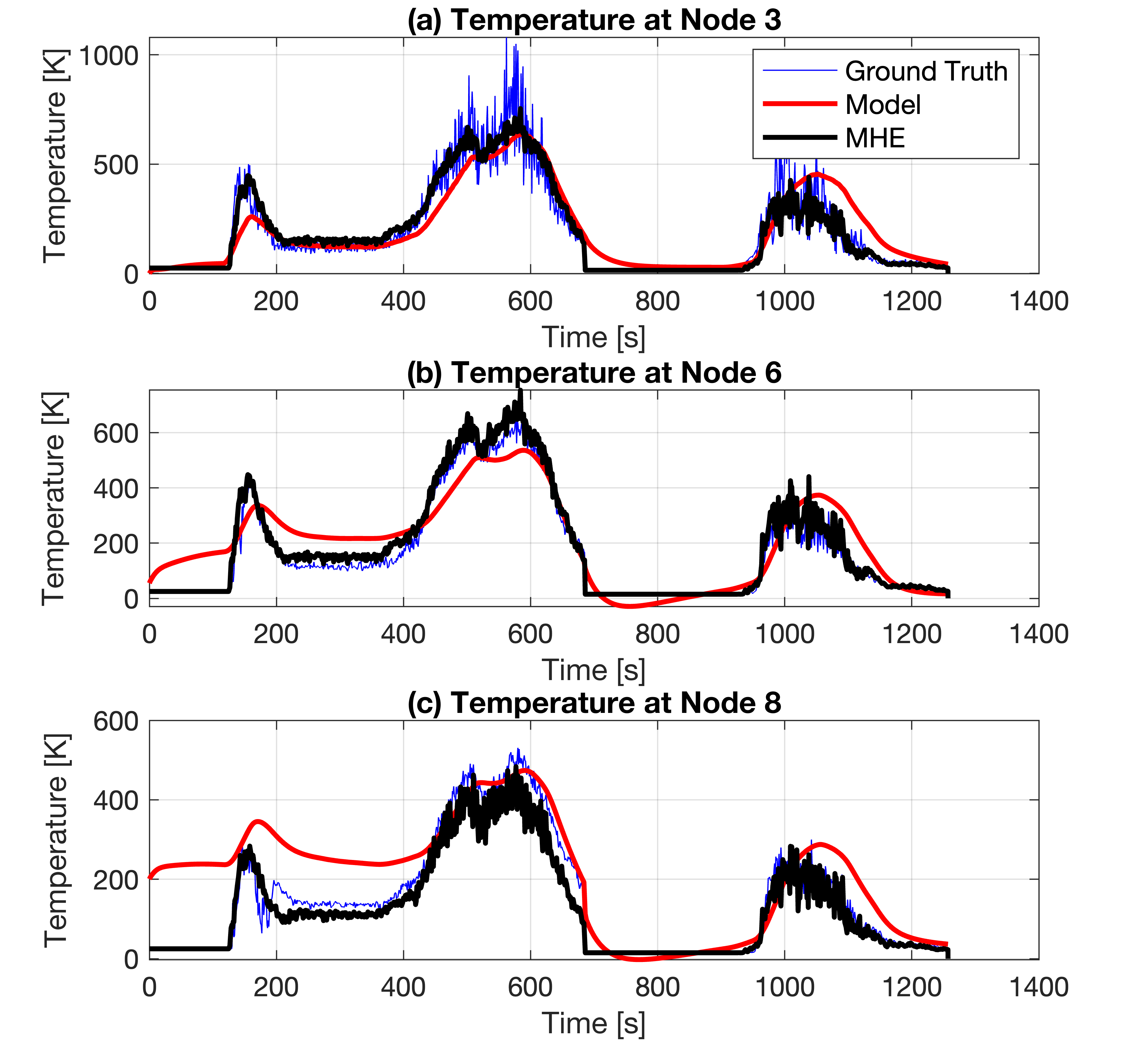}
    \caption{\small Comparison of the thermal reduced order model, ground truth and estimation for validation dataset.(a) Node 3, (b) Node 6, and (c) Node 8, with sensor feedback data from Nodes 1, 5 and 10.}    \label{MHE_node_4_7_9_temp}
\end{figure}

Fig. \ref{MHE_node_4_7_9_temp} shows the thermal estimation performance for some nodes where sensors are not located (nodes 3, 6, and 8), using the aforementioned sensor data. 
It can be seen that the MHE algorithm successfully reconstructs the temperature profiles at the unmeasured node locations, demonstrating the spatial estimation capability of the proposed approach. 
At node locations where sensors are placed (nodes 1, 5, and 10), the MHE estimates closely track the ground truth measurements, with near-perfect alignment at these instrumented nodes. 
Table \ref{table_error_temp} compares the RMSE for the reduced order model and the feedback-based estimation algorithm. 
The estimation errors at sensor locations are found to be negligible, validating the measurement update mechanism within the MHE framework. 
The RMSE analysis provides compelling evidence of the MHE's superiority over reduced order models. 
The open-loop model exhibits RMSE ranging from 28.26 to 173.98 with particularly high errors during the transient phases. 
This indicates that merging reduced order model with real-time sensor feedback can be highly beneficial in improving the accuracy of the state estimation.

\begin{figure}[h!]
    \centering    \includegraphics[width=0.5\textwidth]{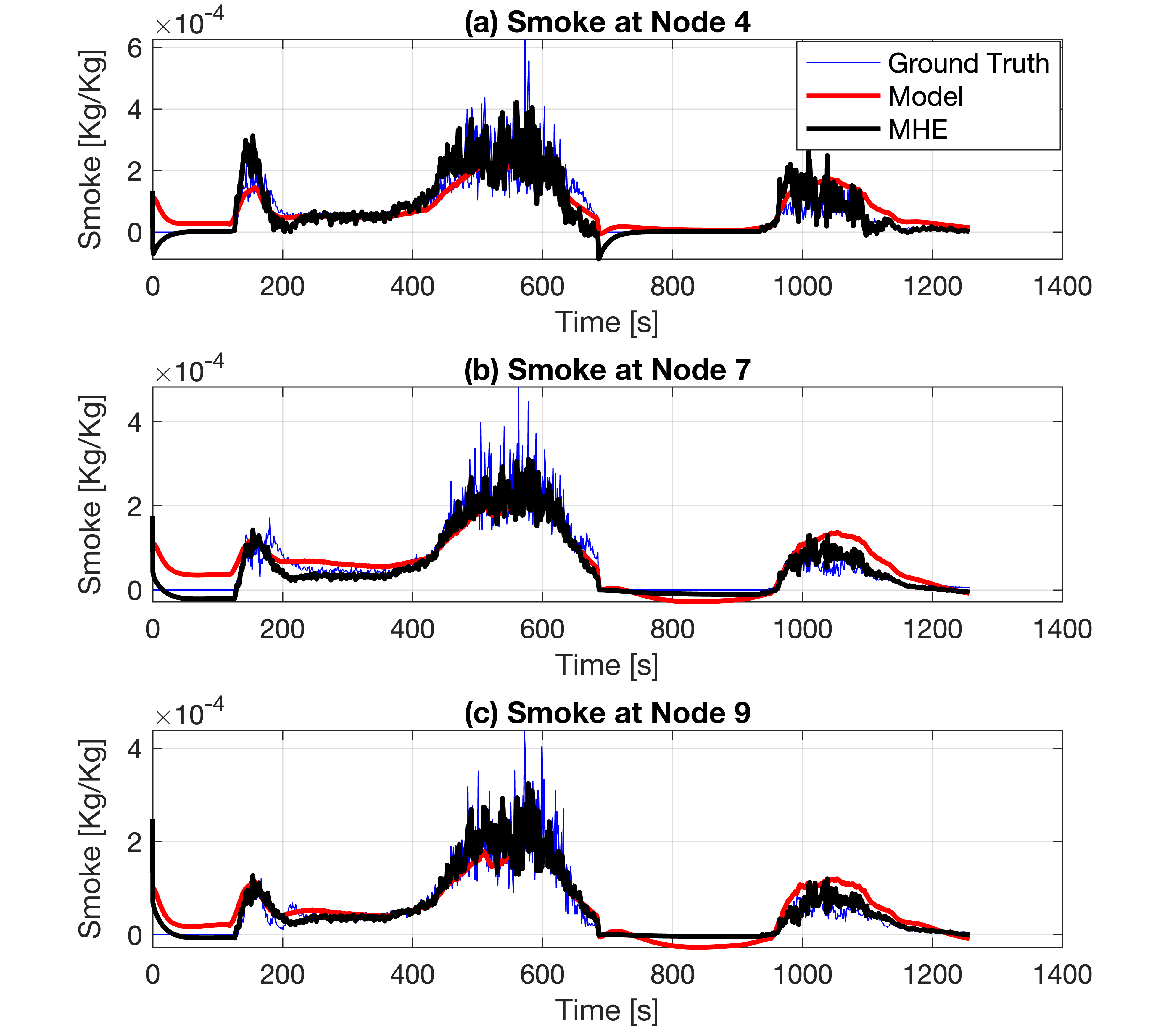}
    \caption{\small Comparison of the smoke reduced order model, ground truth and estimation for validation dataset. (a) Node 4 with using sensor data from Node 5; (b) Node 7 with using sensor data from Node 10; and (c) Node 9 with using sensor data from Node 10.}
    \label{MHE_smoke_fig}
\end{figure}

For smoke estimation, Fig. \ref{MHE_smoke_fig} demonstrates the MHE's capability to reconstruct smoke concentration profiles at locations where no sensors are placed. 
Using sensor data from nodes 1, 5 and 10, the algorithm accurately estimates smoke dynamics at nodes 4, 7, and 9. 
The smoke concentration estimates show good agreement with ground truth data. 
The estimation RMSE for smoke concentration remains within the limit of $1.1757\times10^{-4}$ (as shown in Table \ref{table_error_smoke}), which is particularly notable given the challenging transport smoke dynamics and the sparse sensor configuration. 

\subsection{Robustness analysis with respect to sensor locations}
In this subsection, we evaluate the robustness of the MHE framework under practical sensor deployment constraints. In order to evaluate this, four distinct sensor configurations were analyzed: 
\begin{itemize}
    \item \textit{Case 1:} Sensors at nodes 1 and 5.
    \item \textit{Case 2:} Sensors at node 5.
    \item \textit{Case 3:} Sensors at nodes 5 and 10. 
    \item \textit{Case 4:} Sensors at nodes 1, 5, and 10. 
\end{itemize}

These configurations represent scenarios where sensor placement may be limited by physical constraints, accessibility, or cost considerations in underground mining environments. 
Performance was assessed by comparing the estimate in cases 1 through 3 with the case benchmark 4 that we used to generate the results in the previous subsections.

\begin{figure}[h!]
    \centering    \includegraphics[width=0.5\textwidth]{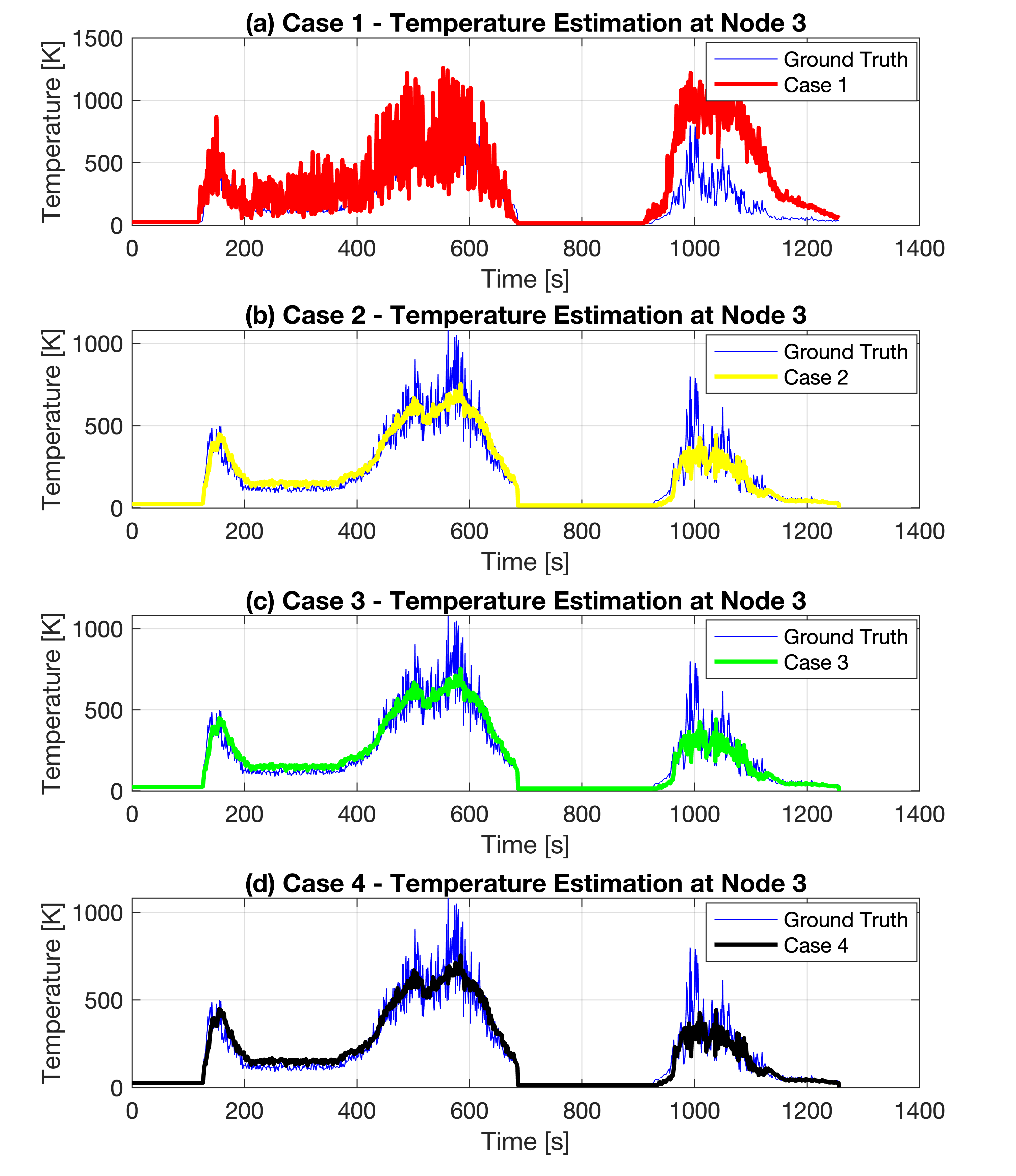}
    \caption{\small Comparison of the thermal ground truth and estimation for validation dataset for Node 3 for (a) Case 1, (b) Case 2, (c) Case 3, (d) Case 4.}
    \label{Robust_temp_4}
\end{figure}

\begin{figure}[h!]
    \centering    \includegraphics[width=0.5\textwidth]{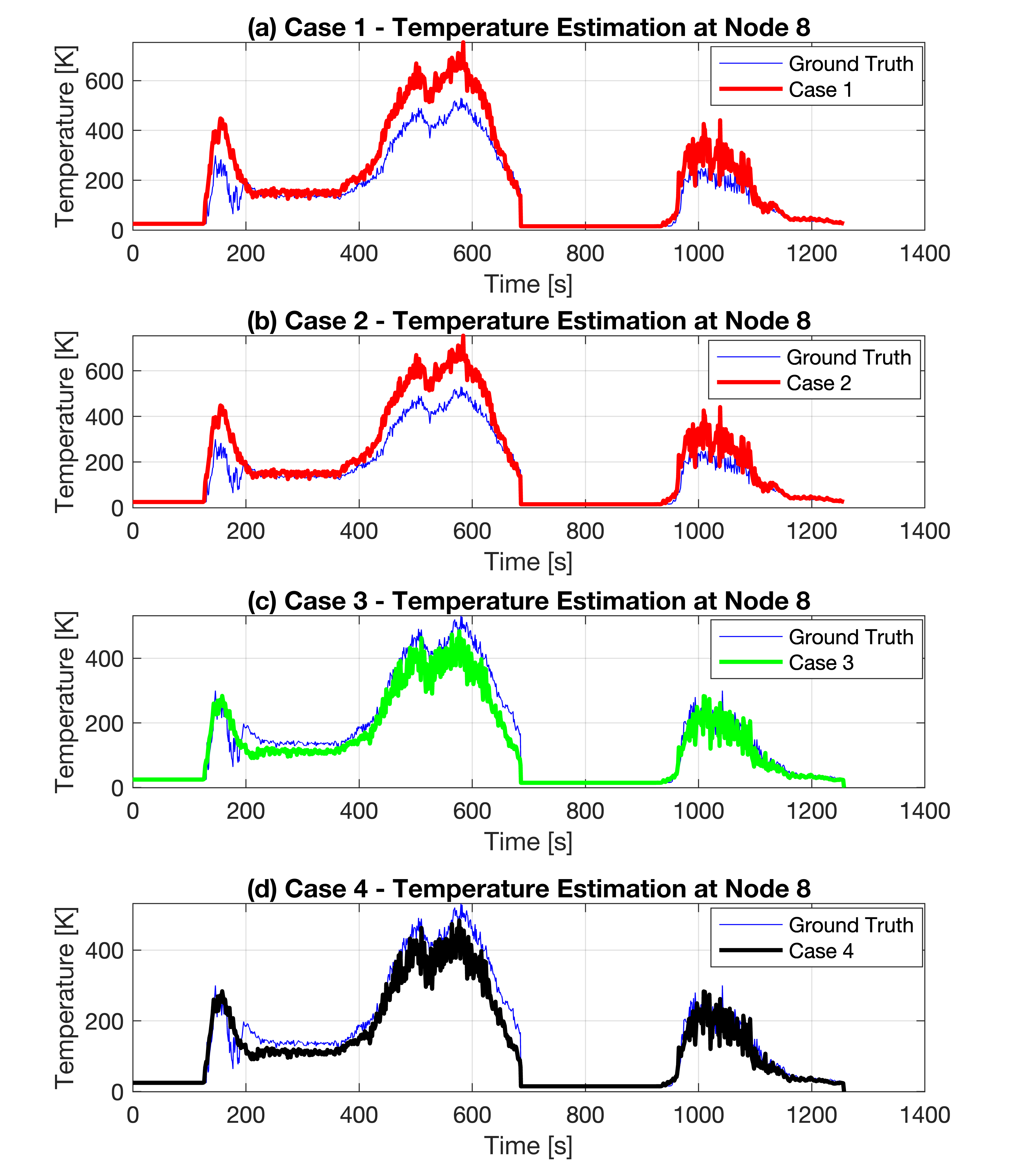}
    \caption{\small Comparison of the thermal ground truth and estimation for validation dataset for Node 8 for (a) Case 1, (b) Case 2, (c) Case 3, (d) Case 4.}
    \label{Robust_temp_9}
\end{figure}

Figs. (\ref{Robust_temp_4}-\ref{Robust_temp_9}) demonstrate the thermal estimation performance for Cases 1 to 4 for node 3 and 8 respectively. 
The estimation framework of case 1 faces greater challenges, and node 3 shows the worst tracking accuracy compared to case 4. 
Similarly, node 8 in case 1, being farther from the sensor locations, exhibits slightly higher estimation errors. 
For case 2, node 3 maintains reasonable accuracy due to its proximity to the measured node, whereas node 8 experiences more significant estimation errors, especially during peak temperature events. 
In case 3, the performance is similar to that in case 4. 
The distributed sensor placement provides better spatial coverage, resulting in improved estimation accuracy at both node 3 and node 8. 
The complementary information from upstream (node 5) and downstream (node 10) sensors enables a more robust state reconstruction throughout the tunnel length.

\begin{figure}[h!]
    \centering
\includegraphics[width=0.5\textwidth]{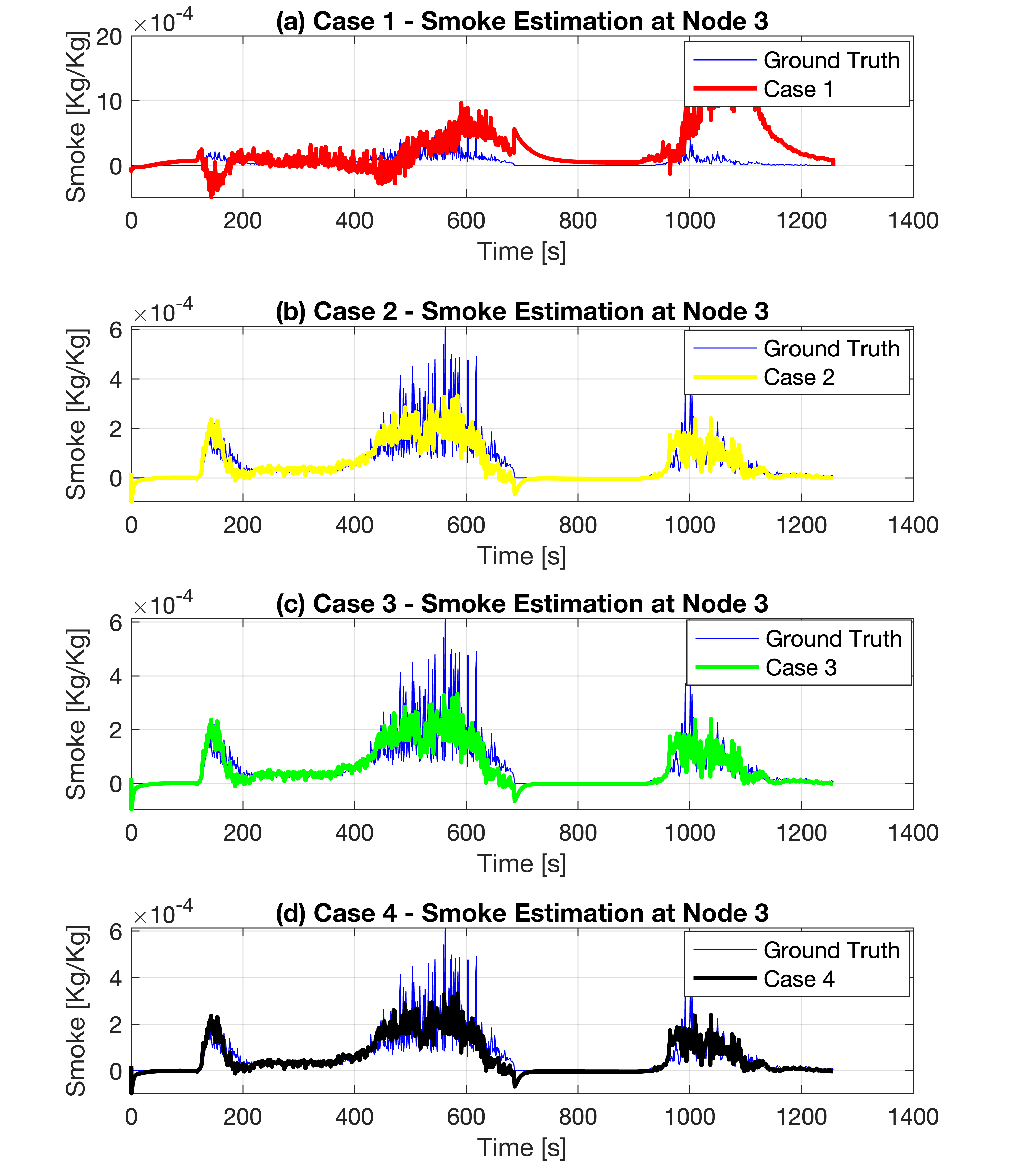}
    \caption{\small Comparison of the smoke ground truth and estimation for validation dataset at Node 3 for (a) Case 1, (b) Case 2, (c) Case 3, (d) Case 4.}
    \label{Robust_smoke_4}
\end{figure}

\begin{figure}[h!]
    \centering
\includegraphics[width=0.5\textwidth]{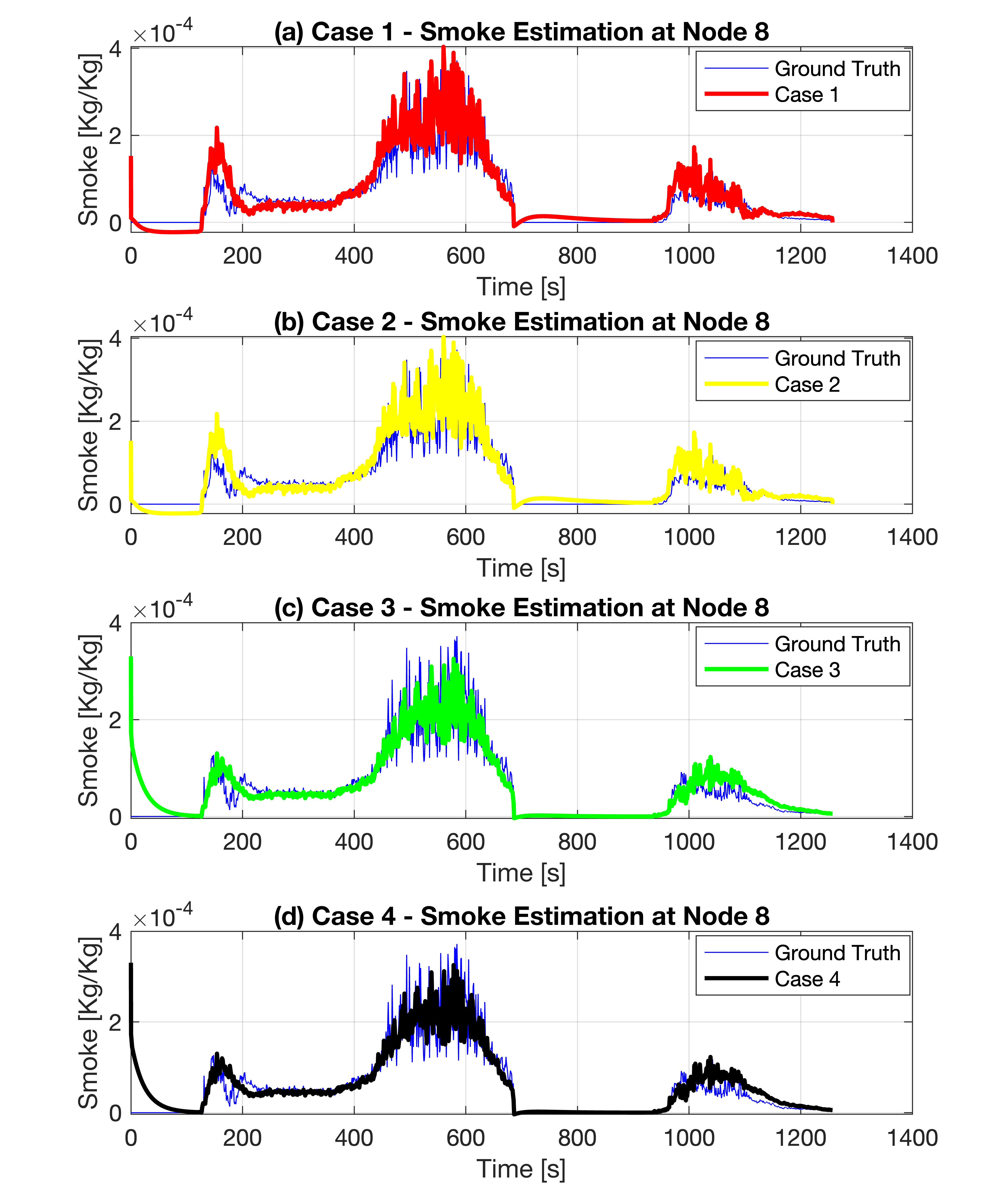}
    \caption{\small Comparison of the smoke ground truth and estimation for validation dataset at Node 9 for (a) Case 1, (b) Case 2, (c) Case 3, (d) Case 4.}
    \label{Robust_smoke_9}
\end{figure}

For smoke dynamics, the comparative analysis across all sensor configurations is summarized in Figs. (\ref{Robust_smoke_4}-\ref{Robust_smoke_9}). Among the test cases, Case 3 and 4 demonstrates superior performance. The MHE algorithm effectively leverages the temporal and smoke history within the moving horizon to compensate for spatial sparsity in measurements, ensuring reliable state estimation for safety-critical monitoring applications.



\section{Conclusion}\label{con}
Large-format lithium-ion batteries are considered promising alternative energy sources for high-power equipment utilized in underground mines.
They effectively eliminate operational hazards such as high-voltage power cables dragged along the mine floor or diesel particulate matter emitted from conventional internal combustion engines. 
However, there have been several incidences of their failure through a rapid combustion process called thermal runaway. This releases a large amount of heat and particulates and mostly noxious or toxic gases. 
Exposure of personnel to these combustion products, especially in the confined spaces of underground mines, can be hazardous. 
High-fidelity models could be developed to mimic the entire combustion event, but they are computationally expensive and cannot be developed for the myriad situations that a mine could encounter. 
They cannot be applied for real-time monitoring of the underground mine environment.

In this work, a cyber-physical systems framework is presented for real-time tracking of temperature and smoke concentration in underground mine tunnels following a thermal runaway event of lithium-ion batteries. 
The proposed approach addresses the computational limitations of high-fidelity CFD models, which are accurate but slow for emergency response. By using data-driven reduced-order models identified from CFD simulations and leveraging real-time sensor feedback, the CPS framework enables rapid spatio-temporal prediction of hazardous conditions using only a sparse set of sensor measurements. The moving horizon estimation algorithm was successfully applied to reconstruct the full state of temperature and smoke dynamics across the entire tunnel, even at locations without sensors. Validation results demonstrated that the MHE-based estimator, which combines reduced order model and sensor feedback, significantly outperformed reduced order models without feedback, with minimal estimation errors at both sensor and non-sensor nodes. Robustness analysis further confirmed that the framework performs reliably under various practical sensor configurations, making it suitable for real-world mining environments where sensor placement may be constrained.

\bibliographystyle{ieeetr}
\bibliography{ref}
\end{document}